\documentclass[useAMS,usenatbib,usedcolumn]{mn2e}
\def\HI{\ifmmode{\rm HI}\else{H\/{\sc i}}\fi}

\def\lsun{\ifmmode{{\mathrm L}_{\odot}}\else{L$_{\odot}$}\fi} 

\def\deg{\hbox{$^\circ$}}

\def\arcsec{\hbox{$^{\prime\prime}$}}

\def\msun{\ifmmode{{\mathrm M}_{\odot}}\else{M$_{\odot}$}\fi} 
\def\msunpc2{\ifmmode{{\mathrm M}_{\odot} \, {\mathrm{pc}}^{-2}}\else{M$_{\odot} \, {\mathrm {pc}}^{-2}$}\fi}

\def\kms{\ifmmode{{\mathrm{km \, s^{-1}}}}\else{${\mathrm{km \, s^{-1}}}$}\fi}

\def\la{\mathrel{\mathchoice {\vcenter{\offinterlineskip\halign{\hfil
$\displaystyle##$\hfil\cr<\cr\sim\cr}}}
{\vcenter{\offinterlineskip\halign{\hfil$\textstyle##$\hfil\cr
<\cr\sim\cr}}}
{\vcenter{\offinterlineskip\halign{\hfil$\scriptstyle##$\hfil\cr
<\cr\sim\cr}}}
{\vcenter{\offinterlineskip\halign{\hfil$\scriptscriptstyle##$\hfil\cr
<\cr\sim\cr}}}}}
\def\ga{\mathrel{\mathchoice {\vcenter{\offinterlineskip\halign{\hfil
$\displaystyle##$\hfil\cr>\cr\sim\cr}}}
{\vcenter{\offinterlineskip\halign{\hfil$\textstyle##$\hfil\cr
>\cr\sim\cr}}}
{\vcenter{\offinterlineskip\halign{\hfil$\scriptstyle##$\hfil\cr
>\cr\sim\cr}}}
{\vcenter{\offinterlineskip\halign{\hfil$\scriptscriptstyle##$\hfil\cr
>\cr\sim\cr}}}}}

\def\aj{AJ}%% Astronomical Journal
\def\apj{ApJ}%% Astrophysical Journal
\def\aap{A\&A}%% Astronomy and Astrophysics
\def\mnras{MNRAS}%% Monthly Notices of the RAS
\def\pasp{PASP}%% Publications of the ASP
\usepackage[figuresright]{rotating}
\usepackage{lscape}
\usepackage{graphics}
\usepackage{epsfig}
\usepackage{multirow}
\usepackage{bigdelim}
\usepackage{bigstrut}

\defcitealias{Noordermeer07b}{N07}

\title[Disk dynamics from IFU data]{Exploring disk galaxy dynamics using
  IFU data}
\author[E.~Noordermeer et al.]
   {E.~Noordermeer,$^1$ 
   M.~R.~Merrifield$^1$\thanks{email: michael.merrifield@nottingham.ac.uk}
   and A.~Arag\'on-Salamanca$^1$ \\ 
   $^1$University of Nottingham, School of Physics and Astronomy, University
       Park, NG7 2RD Nottingham, UK}

\begin{document}

%%\date{v1.2; 26-3-2008}

\maketitle

\begin{abstract}
In order to test the basic equations believed to dictate the dynamics
of disk galaxies, we present and analyze deep two-dimensional spectral
data obtained using the PPAK integral field unit for the early-type
spiral systems NGC~2273, NGC~2985, NGC~3898 and NGC~5533.  We describe
the care needed to obtain and process such data to a point where
reliable kinematic measurements can be obtained from these
observations, and a new more optimal method for deriving the
rotational motion and velocity dispersions in such disk systems.  The
data from NGC~2273 and NGC~2985 show systematic variations in velocity
dispersion with azimuth, as one would expect if the shapes of their
velocity ellipsoids are significantly anisotropic, while the hotter
disks in NGC~3898 and NGC~5533 appear to have fairly isotropic
velocity dispersions.  Correcting the rotational motion for asymmetric
drift using the derived velocity dispersions reproduces the rotation
curves inferred from emission lines reasonably well, implying that
this correction is quite robust, and that the use of the asymmetric
drift equation is valid.  NGC~2985 is sufficiently close to face on
for the data, combined with the asymmetric drift equation, to
determine all three components of the velocity ellipsoid.  The
principal axes of this velocity ellipsoid are found to be in the ratio
$\sigma_z : \sigma_\phi : \sigma_R \approx 0.7 : 0.7 : 1$, which shows
unequivocally that this disk distribution function respects a third
integral of motion.  The ratio is also consistent with the predictions
of epicyclic theory, giving some confidence in the application of this
approximation to even fairly early-type disk galaxies.
\end{abstract}


%%%%%%%%%%%%%%%%%%%%%%%%%%%%%%%%%%%%%%%%%%%%%%%%%%%%%%%%%%%%%%%%%%%%%%%%%%%%%%%
%                                                                             %
%  1. Introduction                                                            %
%  \label{sec:introduction}                                                   %
%                                                                             %
%%%%%%%%%%%%%%%%%%%%%%%%%%%%%%%%%%%%%%%%%%%%%%%%%%%%%%%%%%%%%%%%%%%%%%%%%%%%%%%
\section{Introduction}
\label{sec:introduction}
Disk galaxies are fundamentally dynamical entities, so any
understanding of their structure must pay as much attention to the
velocities of their constituent stars as it does to their positions.
This kinematic information is usually collected through spectroscopy,
where the mean shift and broadening of spectral lines provides
information on the line-of-sight mean velocities and velocity
dispersions of the stellar population that produced the light.  In
the past, such spectra have typically been obtained through long-slit
spectrographs, where a one-dimensional cut through the galaxy is
dispersed in the second dimension on a two-dimensional detector.

Over the years, such data have provided a number of useful insights
into the dynamical properties of these systems.  For example,
observations along the major axes of inclined disk galaxies have
demonstrated that the stellar component rotates more slowly than each
system's circular speed \citep{VanderKruit86, Bottema86}.  This
``asymmetric drift'' is exactly what one expects if the stars also
have significant random velocities, so that the radial structure of
the galaxy is not entirely supported against gravity by its rotational
motion.  Similarly, observations of face-on systems have shown that
the vertical component of the random velocity decreases with radius in
the manner that one might expect if disks are to maintain a constant
scale-height in the presence of radially-decreasing self-gravitation
\citep{VanderKruit86}.

However, the one-dimensional spatial data provided by such long-slit
observations have some serious limitations.  First, these data provide
information on only two dimensions (the spatial one and the
line-of-sight velocity) of the intrinsically six-dimensional phase
space of a galaxy.  This lack of dimensional coverage can be
ameliorated by obtaining long-slit data from a variety of position
angles.  For example, \citet{Gerssen97} obtained data along both the
major and minor axes of the disk galaxy NGC~488 to infer the relative
amplitudes of random motions in two different directions within this
system.  Combining these observations with a physical assumption, such
as the ``epicycle approximation'' which ties random motions in the
radial and azimuthal directions together \citep{Binney87}, then
allowed the solution for the full three-dimensional distribution of
random velocities in the disk of this galaxy.  However, even such
multiple cuts are clearly spatially incomplete and provide a rather
inefficient approach to acquiring kinematic data.  This problem is
compounded by the very faint nature of the outer parts of disk
galaxies, which necessitates long integrations for even a single slit
orientation.

A major step forward came with the development of large-area integral
field units (IFUs), which provide spectral coverage across a full
two-dimensional field of view.  These instruments have revolutionized
our view of elliptical systems through studies such as those
undertaken with the SAURON spectrograph \citep{Emsellem04}.  Results
on the dynamical properties of disk systems have been somewhat slower
to emerge, although some studies are now beginning to bear fruit
\citep[e.g.][]{Westfall08}.  Partly this slower progress arises from
the rather subtler nature of disk galaxy dynamics, as the much smaller
random motions in these systems makes the basic measurements of line
broadening more challenging and prone to systematic error.  Further,
the more complex geometry of these systems, with multiple components,
spiral structure, etc, renders the interpretation of these data a
great deal more difficult.

The potentially-rapid spatial variations in the kinematics of these
systems present particular challenges since the conventional approach
involves averaging together spectra from quite large regions of the
galaxy to obtain a sufficiently high signal-to-noise ratio to obtain
meaningful kinematic measurements.  Such averaging will smooth out any
fine structure in the kinematics and may even introduce significant
biases into the analysis.  For example, \citet{Westfall08} have
recently found that their previous analysis systematically
over-estimated the measurement of some components of random motion
because the spatial averaging over a range of azimuthal angles
combined data with different mean-streaming and random motions,
creating an artificial broadening of the intrinsic velocity
distribution.  Given these difficulties, it is not obvious what the
optimum approach is for deriving even the most basic kinematic
parameters from such two-dimensional data.  For instance, at any given
radius, as one looks around in azimuth away from the major axis, there
is still clearly significant information within the data on the mean
streaming motions of the stars at that radius, but since the
projection of that mean streaming motion into the line of sight varies
with azimuth one cannot simply average the spectra to measure it.

In addition, some of the physical assumptions that have been used in
the interpretation of such kinematic data are somewhat controversial.
The frequently-adopted epicycle approximation, for example, is
strictly only valid when the stars' departures from circular motions
are very small.  \citet{Westfall08} state that their tests show that
this condition is adequately met in the outer parts of all credible
disk galaxies, but \citet{Kuijken94} found that even for the
relatively small amplitude of random motions in a ``cold'' disk
environment like the solar neighbourhood of the Milky Way, this
condition is not sufficiently met, and the resulting error is at the
20 -- 40\% level.  

Even the more generally applicable asymmetric drift equation can only
be used if a number of assumptions are made.  It is, for example,
often assumed that the random motions decrease smoothly and
exponentially with radius, to force the derivatives of these
quantities to be well behaved.  Although this assumption is supported
by observation of a number of system
\citep[e.g.][]{VanderKruit86, Bottema93, Gerssen97} including
our own galaxy \citep{Lewis89}, it is now known that in some cases the
random motions do not vary with radius in this simple manner in disk
systems \citep[e.g.][]{Merrett06, Noordermeer08}.  The applicability
of this equation has broader implications, since the correction for
asymmetric drift has been used to estimate the circular speeds of S0
galaxies for which no emission-line rotation curves exist
\citep*{Bedregal06b}.  These circular speeds have then been used to
study the Tully--Fisher relation for these systems to determine
whether they are simply faded spiral galaxies \citep*{Bedregal06a}.
If it turns out that the asymmetric drift correction does not provide
a robust way to estimate circular speeds for these fairly ``hot''
disks, then such techniques for inferring the life histories of S0
systems are brought into serious question.

In this paper, we revisit all these issues with a view to creating and
testing a more robust framework for analyzing the kinematics of disk
galaxies, and testing the assumptions that go into analyzing their
dynamics.  In Section~\ref{sec:sample+obs}, we describe the selection
of a sample of four fairly early-type disk galaxies.  As such, their
observed kinematics are not contaminated by strong localized spiral
features, so we can restrict this first analysis of disk galaxy
dynamics to their more global properties.  In addition, their early
types mean that they are directly comparable to S0 systems, allowing
us to test the validity of asymmetric drift corrections for such
galaxies.  However, they are all also chosen to display good
emission-line rotation curves, so that we have direct comparators for
the dynamically-inferred stellar rotation curves.
Section~\ref{sec:sample+obs} also describes the rather careful
observations undertaken using the PPAK IFU on the Calar Alto 3.5m
Telescope to minimize systematic errors in the requisite high-quality
spectral data, and Section~\ref{sec:reduction} presents the manner in
which these data were reduced while taking care to preserve the
dynamical signal in an uncompromised form.
Section~\ref{sec:kinematics} introduces a new technique for extracting
in a more optimal manner the mean streaming velocity and random
motions from such two-dimensional data, and presents the resulting
inferred kinematics.  Section~\ref{sec:dynamics} explores the dynamics
that can be derived from these data, discussing the shape of the
velocity ellipsoid, and the validity of both the asymmetric drift
equation and the epicycle approximation.  Finally, we draw some
conclusions in Section~\ref{sec:conclusions}.

%%%%%%%%%%%%%%%%%%%%%%%%%%%%%%%%%%%%%%%%%%%%%%%%%%%%%%%%%%%%%%%%%%%%%%%%%%%%%%%
%                                                                             %
%  2. Sample selection and observations                                       %
%  \label{sec:sample+obs}                                                     %
%                                                                             %
%%%%%%%%%%%%%%%%%%%%%%%%%%%%%%%%%%%%%%%%%%%%%%%%%%%%%%%%%%%%%%%%%%%%%%%%%%%%%%%
\section{Sample selection and observations}
\label{sec:sample+obs}

%%%%%%%%%%%%%%%%%%%%%%%%%%%%%%%%%%%%%%%%%%%%%%%%%%%%%%%%%%%%%%%%%%%%%%%%%%%%%%%
%                                                                             %
% BEGIN TABLE 1: BASIC DATA                                                   %
% label: {table:sample}                                                       %
%%%%%%%%%%%%%%%%%%%%%%%%%%%%%%%%%%%%%%%%%%%%%%%%%%%%%%%%%%%%%%%%%%%%%%%%%%%%%%%
\begin{table*}
 \begin{minipage}{12.3cm}
  \centering
   \caption[Basic data for sample galaxies]
   {Sample galaxies: basic data\label{table:sample}}
  
   \begin{tabular}{llccccccc}
    \hline 
    \multicolumn{1}{c}{Name} & \multicolumn{1}{c}{Type} & D & 
    L$_{\mathrm R}$ & $\mu_{0,R}$ & $B-R$ & $v_{\mathrm {max}}$ & 
    $v_{\mathrm {asymp}}$ & i \\     
    
     & & Mpc & $10^{10}$ L$_\odot$ & $\frac{\mathrm{mag}}{\mathrm{arcsec}^2}$
    & mag & \kms & \kms & \deg \\    
   
    \multicolumn{1}{c}{(1)} & \multicolumn{1}{c}{(2)} & (3) & (4) & (5) & (6)
    & (7) & (8) & (9) \\   
    \hline 

    NGC 2273 & SB(r)a       & 27.3 & 1.84  & 15.84 & 1.5 & 260 & 190 & 55 \\
    NGC 2985 & (R')SA(rs)ab & 21.1 & 3.05  & 15.56 & 1.1 & 255 & 220 & 37 \\
    NGC 3898 & SA(s)ab      & 18.9 & 1.72  & 15.35 & 1.3 & 270 & 250 & 69 \\
    NGC 5533 & SA(rs)ab     & 54.3 & 5.92  & 16.27 & 1.4 & 300 & 230 & 53 \\
    \hline
  \end{tabular}
 \end{minipage}
\vskip 0.1cm
\begin{minipage}{12.3cm}
{\it Notes} -- Col.\ (1), name; Col.\ (2), morphological type; Col.\
(3), distance (based on Hubble flow, corrected for Virgo-centric
inflow and assuming h=0.75); Cols\ (4)~and (5), R-band total
luminosity and central surface brightness(corrected for Galactic
foreground extinction); Col.\ (6) $B-R$ colour; Cols.\ (7) and (8),
maximum and asymptotic rotation velocities; Col.\ (9)
inclination. Col.\ (2) from NASA Extragalactic Database; Col.\ (3)
from \citet{Noordermeer05}; Cols (4) -- (6) from
\citet{Noordermeer07a}; and Cols (7) -- (9) from
\citetalias{Noordermeer07b}.
\end{minipage}
\end{table*}  
%%%%%%%%%%%%%%%%%%%%%%%%%%%%%%%%%%%%%%%%%%%%%%%%%%%%%%%%%%%%%%%%%%%%%%%%%%%%%%%
%                                                                             %
% END TABLE 1: BASIC DATA                                                     %
%                                                                             %
%%%%%%%%%%%%%%%%%%%%%%%%%%%%%%%%%%%%%%%%%%%%%%%%%%%%%%%%%%%%%%%%%%%%%%%%%%%%%%%

\subsection{Sample selection}
\label{subsec:sample}
Since the intention of this study is to investigate the viability of
measuring the large-scale stellar dynamics of disk systems, we have
targeted early-type disk galaxies that do not contain significant
confusing spiral structure.  In addition, these systems are
sufficiently early in the Hubble sequence to be comparable to S0
galaxies, since we are seeking to ascertain whether one can correct
for asymmetric drift in such potentially-hot disks.  However, we also
require sufficient gas in each system for there to be a reliable
emission-line rotation curve for comparison, so we have selected
galaxies from the gas-rich early-type disk galaxy sample of
\citet[][hereafter N07]{Noordermeer07b}, for which high-quality
rotation curves have already been derived using \HI\ observations and
long-slit optical emission line spectra.

In addition, we restricted the selection to galaxies that:
\begin{enumerate}
\item are visible from Calar Alto at the time of observation;
\item are large enough on the sky for us to co-add many spectra to
  obtain adequate signal;
\item are small enough on the sky to fit most of the light of the
  galaxy within two pointings of the PPAK IFU;
\item are inclined at an inclination of $i \ga 40\deg$ in order to
  ensure that rotational motion and other in-plane velocities are
  observable in the line-of-sight velocities;
\item are inclined at an inclination of $i \la 70\deg$ in order to
  resolve the minor axis of the system and to avoid extreme projection
  effects along the line of sight.
\end{enumerate}
The basic properties of the four galaxies in this study, selected by
these criteria, are presented in Table~\ref{table:sample}.

\subsection{Observations}
\label{subsec:observations}
The galaxies in the sample were observed with the fibre-based
integral-field unit PPAK, mounted on the PMAS spectrograph of the
German--Spanish 3.5m telescope at Calar Alto Observatory.  The PPAK
unit comprises 331 fibres, each with a diameter on the sky of
2.7\arcsec, packed in a hexagonal grid with minimum and maximum
diameters of 64 and 74\arcsec\ respectively; the large field of view
of this IFU makes it very well-suited to studies of nearby galaxies.
An additional 36 fibres (6 groups of 6 fibres each) are placed at a
distance of 72\arcsec\ from the IFU centre and can, as long as they
are not polluted by light from the main object, be used to sample the
surrounding sky.  Finally, 15 fibres are not illuminated from the sky,
but are connected to a calibration unit.  These calibration fibres can
be illuminated with light from lamps during the science exposures,
allowing synchronous spectral calibration, which is important in
correcting for flexure between exposures, to prevent such systematic
problems from compromising the measured velocity dispersions.  A more
detailed description of the PPAK fibre-unit is provided by
\citet{Kelz06}.

The PPAK unit feeds the spectra from all fibres into the PMAS
spectrograph.  To achieve the desired spectral resolution, we used the
J1200 grating in second order and in `backward blaze' \citep{Roth05,
Kelz06}.  By rotating the grating appropriately, a spectral window of
4950\AA -- 5400\AA\ was selected, covering the Mg~b 5175\AA\
absorption lines and many surrounding weaker iron features.  The
spectral resolution with this setup is about 8200, corresponding to a
FWHM wavelength resolution at our central wavelength of 0.63\AA.  In
practice, the spectral resolution was found to vary systematically
with position on the detector chip, presumably caused by a slight tilt
of the detector in the focal plane, and the average effective
resolution as measured from our final spectra was approximately
0.8\AA\ (FWHM), corresponding to an instrumental velocity dispersion
of 20~\kms.  As we describe below, these variations in resolution with
position were dealt with in the data reduction to prevent the
introduction of systematic biases in the measured velocity
dispersions.

The observations were carried out under photometric conditions on the
nights 2007 January 17 -- 20.  Even with the large size of the PPAK
IFU, the target galaxies were too large to fit in the field of view of
the fibre bundle, so each system was observed in two pointings.  A few
short exposures, giving a total integration times of 1 -- 1.5 hours,
were taken of the bright, central regions of each galaxy, while
multiple longer exposures, giving total exposure times between 4.5 and
6 hours, were taken at fields along the major axis, offset such that
they had a small overlap with the central pointing.  A summary of
these observations is given in Table~\ref{table:observations}.
%%%%%%%%%%%%%%%%%%%%%%%%%%%%%%%%%%%%%%%%%%%%%%%%%%%%%%%%%%%%%%%%%%%%%%%%%%%%%%%
%                                                                             %
% BEGIN TABLE 2: OBSERVATIONS                                                 %
% label: {table:observations}                                                 %
%%%%%%%%%%%%%%%%%%%%%%%%%%%%%%%%%%%%%%%%%%%%%%%%%%%%%%%%%%%%%%%%%%%%%%%%%%%%%%%
\begin{table*}
 \begin{minipage}{13.cm}
  \centering
   \caption[Observations]
   {Summary of the observations \label{table:observations}}

   \begin{tabular}{llclcc}
    \hline
    
    galaxy & pointing & observing dates & total exposure time & 
             average seeing & photometric?                           \\
           &          & January 2007    &   \hspace{0.8cm}s   & 
             \arcsec        &                                        \\
    \hline

    NGC 2273 & centre     & 17     & \hspace{0.5cm}3600  & 1.8 & yes \\
             & south-west & 17--20 & \hspace{0.5cm}18000 & 1.3 & yes \\
    NGC 2985 & centre     & 17     & \hspace{0.5cm}3600  & 1.6 & yes \\
             & south      & 18--20 & \hspace{0.5cm}20400 & 1.1 & yes \\
    NGC 3898 & centre     & 17     & \hspace{0.5cm}3600  & 1.5 & yes \\
             & north-west & 17--20 & \hspace{0.5cm}21600 & 1.2 & yes \\
    NGC 5533 & centre     & 17     & \hspace{0.5cm}5400  & 1.4 & yes \\
             & south-west & 18--20 & \hspace{0.5cm}16500 & 1.1 & yes \\
    \hline
   \end{tabular}
 \end{minipage}
\end{table*}  
%%%%%%%%%%%%%%%%%%%%%%%%%%%%%%%%%%%%%%%%%%%%%%%%%%%%%%%%%%%%%%%%%%%%%%%%%%%%%%%
%                                                                             %
% END TABLE 2: OBSERVATIONS                                                   %
%                                                                             %
%%%%%%%%%%%%%%%%%%%%%%%%%%%%%%%%%%%%%%%%%%%%%%%%%%%%%%%%%%%%%%%%%%%%%%%%%%%%%%%

Since the galaxies are large, the IFU's sky fibres were often
contaminated by light from the galaxy, and so would not have been
suitable for sky subtraction.  We therefore obtained short (300
second) exposures of blank fields before and after the galaxy
exposures; by averaging the spectra from these exposures in the fibres
that contained no light from foreground stars, high signal-to-noise
ratio sky spectra were obtained (see
Section~\ref{subsec:skysubtraction}).  Further calibration of all
these on-sky exposures was obtained by briefly (for typically 5
seconds) illuminating the 15 calibration fibres with a thorium--argon
(ThAr) arc lamp, in order to correct for any intra-exposure flexure of
the instrument (see Section~\ref{subsec:flexcorr}).

\begin{figure}
 \centerline{
   \psfig{figure=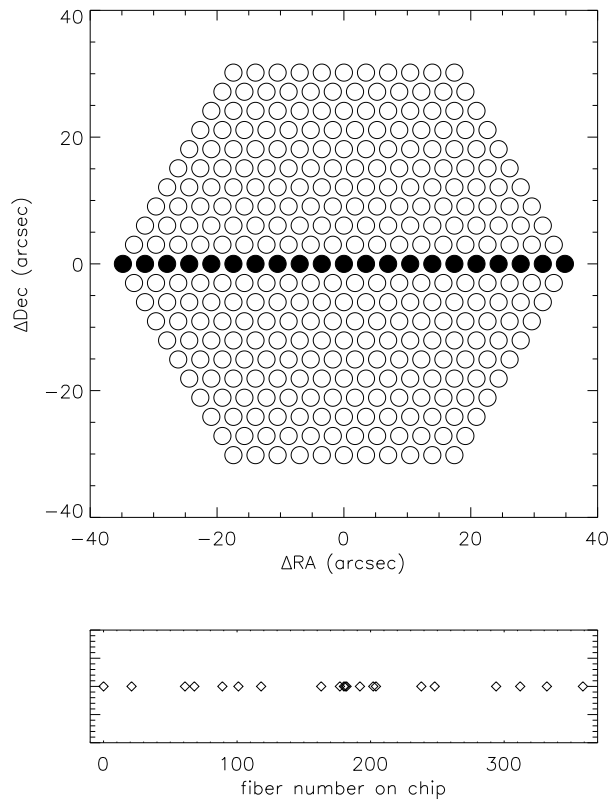,height=10.5cm}
            }
  \caption{Illustration of the drift-scan procedure used for the
   observation of the template star HD~20893. The top panel shows the
   illuminated fibres: the telescope was drifted such that the star
   spectrum is observed in each fibre in the central row. The bottom
   panel shows the location of the illuminated spectra, which span the
   detector.\label{fig:templatestarscan}}
\end{figure}

During the run, a total of 13 velocity template stars were observed
(see Table~\ref{table:templatestars}).  Most stars are of late
spectral type (G and K), but an A star and two F stars were also
included to ensure that we could construct composite spectra that
matched the galaxies.  In order to quantify the variation of the
spectral resolution with position on the chip (see above), we drifted
the telescope in the east--west direction during these exposures, such
that we obtained the star spectrum in each fibre along a central row
in the IFU (see Figure~\ref{fig:templatestarscan}, upper panel).  Due
to the ordering of fibres in the spectrograph, this sampling also spanned
the spectrograph detector (see Figure~\ref{fig:templatestarscan},
lower panel), allowing us to calibrate out the variations in spectral
resolution with position that would otherwise have compromised our
ability to measure velocity dispersions with the necessary accuracy.

%%%%%%%%%%%%%%%%%%%%%%%%%%%%%%%%%%%%%%%%%%%%%%%%%%%%%%%%%%%%%%%%%%%%%%%%%%%%%%%
%                                                                             %
% BEGIN TABLE 3: TEMPLATE STARS                                               %
% label: {table:templatestars}                                                %
%%%%%%%%%%%%%%%%%%%%%%%%%%%%%%%%%%%%%%%%%%%%%%%%%%%%%%%%%%%%%%%%%%%%%%%%%%%%%%%
\begin{table}
 \centering
  \caption[Template stars]
  {Observed velocity template stars \label{table:templatestars}}
  \begin{tabular}{ll}
   \hline
   star name & spectral type \\
   \hline
   HD 65900  & A1 V       \\
   HD 137391 & F0 V       \\
   HD 136202 & F8 III--IV \\
   HD 139641 & G7.5 IIIb  \\
   HD 38656  & G8 III     \\
   HD 41636  & G9 III     \\
   HD 28946  & K0         \\
   HD 135482 & K0 III     \\
   HD 139195 & K0p        \\
   HD 23841  & K1 III     \\
   HD 48433  & K1 III     \\
   HD 20893  & K3 III     \\
   HD 102328 & K3 III     \\
   \hline
  \end{tabular}
\end{table}  
%%%%%%%%%%%%%%%%%%%%%%%%%%%%%%%%%%%%%%%%%%%%%%%%%%%%%%%%%%%%%%%%%%%%%%%%%%%%%%%
%                                                                             %
% END TABLE 3: TEMPLATE STARS                                                 %
%                                                                             %
%%%%%%%%%%%%%%%%%%%%%%%%%%%%%%%%%%%%%%%%%%%%%%%%%%%%%%%%%%%%%%%%%%%%%%%%%%%%%%%

Finally, in addition to standard bias frames, two sets of calibration
images were taken at the beginning and end of each observing night.
First, a series of dome flat exposures was taken using a conventional
incandescent lamp.  The calibration fibres were simultaneously
illuminated with a halogen light-source.  The resulting spectra are
bright and featureless and are used to define and trace the apertures
for the individual fibres over the detector, as well as to correct for
fibre-to-fibre throughput variations (see
Section~\ref{subsec:specextract}).  Second, additional dome flats were
obtained using a bright arc lamp, simultaneously illuminating the
calibration fibres with the same ThAr lamp as used during the
night-time, on-sky exposures.  These frames are used to determine the
wavelength solution (see Section~\ref{subsec:specextract}).

%%%%%%%%%%%%%%%%%%%%%%%%%%%%%%%%%%%%%%%%%%%%%%%%%%%%%%%%%%%%%%%%%%%%%%%%%%%%%%%
%                                                                             %
%  3. Data reduction                                                          %
%  \label{sec:reduction }                                                     %
%                                                                             %
%%%%%%%%%%%%%%%%%%%%%%%%%%%%%%%%%%%%%%%%%%%%%%%%%%%%%%%%%%%%%%%%%%%%%%%%%%%%%%%
\section{Data reduction}
\label{sec:reduction}

\subsection{Initial steps}
\label{subsec:initsteps}
The initial stages in the data reduction were performed within the
IRAF environment.\footnote{IRAF is distributed by the National Optical
Astronomy Observatories, which are operated by the Association of
Universities for Research in Astronomy, Inc., under cooperative
agreement with the National Science Foundation.}  The readout bias in
each raw image was subtracted using the overscan region of the chip.
The remaining structure in the noise was removed using standard bias
frames.  

To correct for pixel-to-pixel sensitivity variations, we used a set of
special dome flats, kindly provided by Marc Verheijen, taken with the
spectrograph so far out of focus that the emission from the individual
fibres is smeared out and fills in the inter-fibre regions on the
chip.  From these frames, a response image was created by fitting a
low-order polynomial in the spectral direction and calculating the
ratio between the raw image and the fit, thus removing the large-scale
spectral shape and the intensity variations between fibre- and
inter-fibre-regions from the original images.  The individual
bias-subtracted images were then divided by this response image.

Bad pixels were identified in the combined bias image, by finding
individual pixels as well as entire columns and rows that deviate
strongly from their neighbours.  Once the bad regions were identified,
the corresponding pixels in all other images were corrected by linear
interpolation from their neighbours.  Finally, an initial pass was
made to remove cosmic rays from individual frames using the
`L.A.Cosmic' procedure \citep{VanDokkum01}.

\subsection{Corrections for flexure}
\label{subsec:flexcorr}
PMAS is known to suffer significantly from flexure, particularly when
objects are past the meridian and starting to set \citep{Roth05,
Verheijen04}.  To mitigate this problem, we only observed galaxies
while they were rising.  However, it was still important to monitor
and correct for any residual flexure.  In principle, one could use the
ThAr spectra in the 15 calibration fibres for each exposure
directly to derive individual distortion corrections and wavelength
solutions.  However, due to the low signal-to-noise ratio of these
spectra, the small number of ThAr lines in our wavelength range and
the sparse sampling in the spatial direction (15 calibration fibres
for a total of 367 sky spectra), this approach is in practice not
viable.  Instead, for each on-sky exposure, we compared the positions
of a few well-detected emission lines in the calibration fibres with
those in the averaged arc-lamp dome flat exposure.  The average
difference between the positions on both images then gives the linear
flexure of the instrument between the dome flats and the on-sky
exposure.  In practice, these shifts between exposures were found to
be only of the order of one pixel or less, and were effectively
removed by this procedure.  The small size of the effect means that
shifts in the wavelength solution during individual exposures would
not blur the spectra sufficiently to compromise our ability to
determine velocity dispersions accurately.

Once the shift in the calibration ThAr lines between each on-sky
exposure and the arc dome flat had been determined, new arc and dome
flat images were created, shifted to match the on-sky images.  This
approach is preferred to shifting the on-sky data since the dome flat
images have much higher signal-to-noise ratios, so are not compromised
by this sub-pixel interpolated shift.  Thus, each on-sky image ends up
with its own calibration images, corrected for any flexure present at
its particular telescope orientation.

\subsection{Extraction and calibration of the spectra}
\label{subsec:specextract}
The HYDRA package in IRAF was used to extract and calibrate the
spectra from the individual fibres.  First, the aperture for each
fibre was determined using the continuum-lamp dome flat (corrected for
flexure as described above).  An identification table was used to
distinguish between object, sky and calibration fibres.  The
continuum-lamp spectra were traced along the spectral direction, and a
function fitted to describe the fibre positions on the chip.  The
dome flat spectra were then extracted and their relative intensities
used to determine the fibre-to-fibre throughput variations.
Similarly, the arc-lamp spectra (corrected for flexure) were extracted
using the same positional fit, the lines in the spectra were
identified, and a solution produced by fitting a low-order polynomial
to describe wavelength as a function of position in the spectrum.  
Finally, the object spectra were extracted with the same spatial
trace, and were re-binned to a logarithmic wavelength scale based on
the derived spectral calibration.

To assess independently the accuracy of these wavelength calibrations,
we measured the wavelengths of a number of sky lines in a selection of
the final spectra.  We found excellent agreement with the literature
wavelengths, with a negligible mean offset and an RMS scatter of
0.05\AA, corresponding to an insignificant velocity error of 3~\kms.

\subsection{Sky subtraction}
\label{subsec:skysubtraction}
For the subtraction of the contribution of the sky to our template
star spectra, we averaged the spectra from all fibres which were not
illuminated by the star.  For the galaxy spectra, we averaged all
spectra from the corresponding separate sky exposure (again rejecting
any fibres which were contaminated by foreground stars) and rescaled
according to the exposure time of the latter.  The resulting sky
spectra were then subtracted from each individual spectrum to yield a
set of cleaned and fully calibrated spectra.

\subsection{Final steps}
\label{subsec:finalsteps}
The final step in the data reduction process was to combine the
individual exposures of the same fields to increase the
signal-to-noise ratio and to reject residual cosmic rays and bad
pixels that had not been removed in the previous steps.  Thus, the end
result of the reductions are, for each galaxy, two sets of fully
cleaned and calibrated spectra, one for each pointing.

To facilitate this step, we used the `Euro3D Visualisation Tool'
\citep{Sanchez04b} to display the reduced spectral and spatial data as
a cube and to visually inspect features in the spectra.  The software
requires as additional input a table describing the position of the
331 fibres in the central hexagon on the sky, which was kindly
provided to us by the Calar Alto staff.  This tool also allowed us to
make interpolated 2D images of the integrated continuum emission in
each pointing.  The images for the central pointings were then
registered to the R-band images from \citet{Noordermeer07a} to
accurately determine the centre of the galaxies in all these
pointings, thus allowing the creation of a single registered stacked
spectral data cube.

\begin{figure*}
 \centerline{\psfig{figure=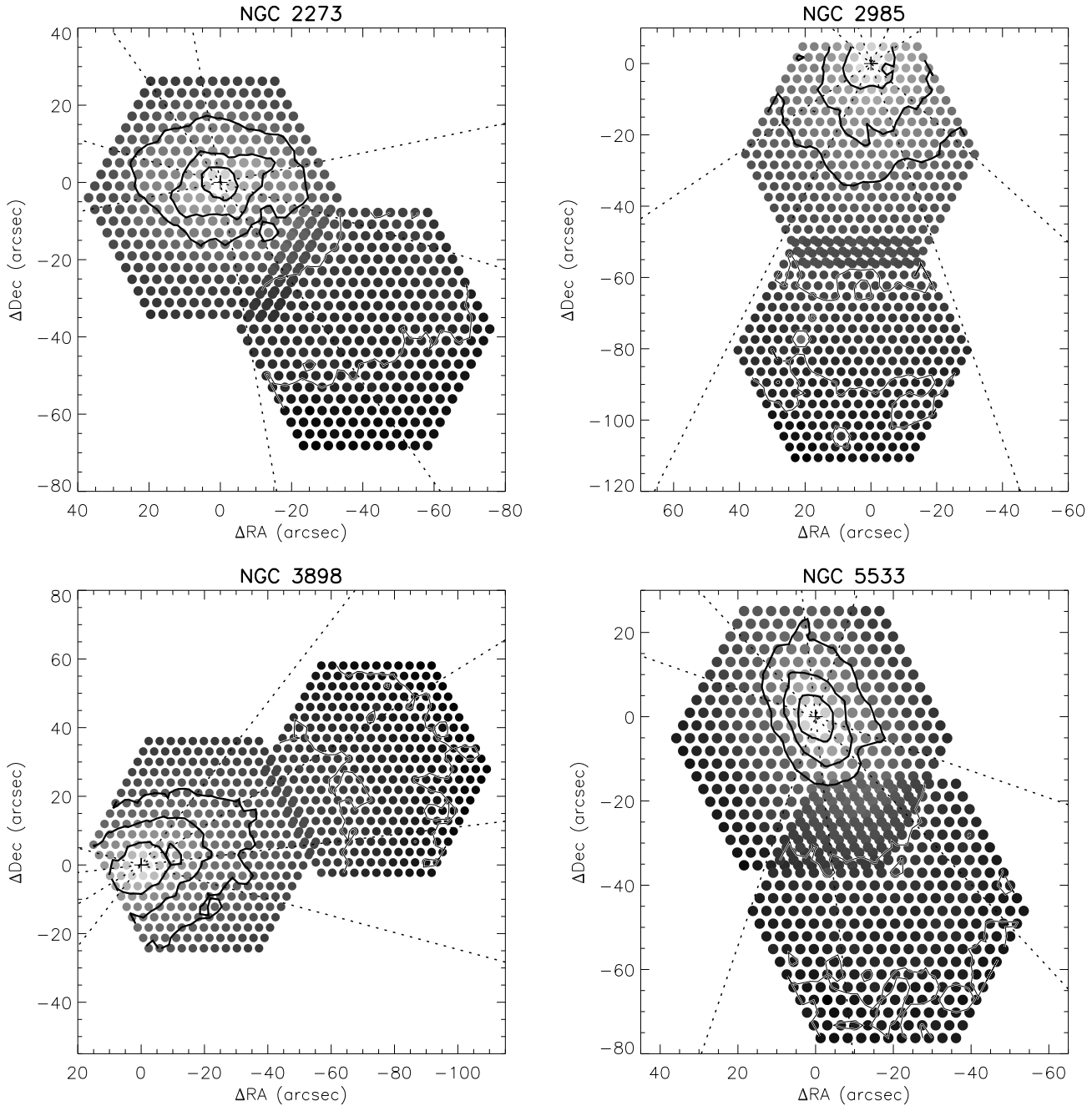,height=17.5cm}}
  \caption{Reconstructed images of total intensity for the four
    galaxies in the sample. The dots indicate the brightness of the
    integrated emission in each fibre, on a logarithmic
    gray-scale. The sizes of the dots are roughly equal to the true
    size of the fibres on the sky.  The thick black contours show
    constant signal-to-noise ratios (per 0.21\AA\ pixel) of 20, 10 and
    5 (from the centre outwards) in the central pointings, while the
    thin black contours indicate signal-to-noise ratios of 5 and 2 in
    the outer pointings. The locations of the centres of the galaxies
    are indicated with the plus-signs. Dotted lines indicate the
    regions within azimuthal angles of 30\deg\ from the major and the
    minor axes. \label{fig:totint}}
\end{figure*}

For the off-centre pointings of the galaxies, there was generally too
little structure in the continuum brightness distribution to register
accurately to the R-band images, so for these we used the recorded
telescope offsets between the central and the off-centre pointings
instead.  Using those cases where there was an appropriate calibrator
(such as a foreground star), we found that the error in these offsets was
significantly smaller than a fibre diameter, and is entirely adequate
for the purposes of measuring the large-scale dynamics of these
systems.  Figure~\ref{fig:totint} shows the resulting spatial coverage
of the pair of pointing on each galaxy, and the reconstructed image
that one obtains by collapsing the spectra into an integrated light
measurement.

%%%%%%%%%%%%%%%%%%%%%%%%%%%%%%%%%%%%%%%%%%%%%%%%%%%%%%%%%%%%%%%%%%%%%%%%%%%%%%%
%                                                                             %
%  3. Analysis                                                                %
%  \label{sec:kiematics}                                                      %
%                                                                             %
%%%%%%%%%%%%%%%%%%%%%%%%%%%%%%%%%%%%%%%%%%%%%%%%%%%%%%%%%%%%%%%%%%%%%%%%%%%%%%%
\section{Inferring the kinematics}
\label{sec:kinematics}
Having produced a set of fully calibrated spectra, we are now in a
position to infer each galaxy's kinematic properties.  As highlighted
in the Introduction, the issues in studies of disk galaxies are rather
different from those in elliptical systems for two reasons.  First,
the signals are rather subtler in that the random motions, although
critical to understanding the dynamics of these systems, are
relatively small so must be extracted with care.  Second, the
observable kinematics can vary quite rapidly with position, so one
must also be careful when averaging together data to obtain adequate
signal-to-noise ratios that one does not wash out these signals in a
destructive manner.  To deal with these issues, we have adapted some
existing software and developed some new algorithms, as we now
describe.

\subsection{Fitting technique}
\label{subsec:PPXF}
Before getting to the details of how we combine data to obtain the
requisite signal-to-noise ratio spectra, we must first choose an
appropriate method for extracting kinematic parameters and errors from
such spectra, since this part of the process will inform us as to the
signal-to-noise ratios required in the combined data.  After some
experimentation, we adopted the Penalized Pixel Fitting Method (PPXF)
from \citet{Cappellari04} to measure the velocity shift and
dispersion, since it proved robust and effective.  For each spectrum,
this method finds the linear combination of velocity template star
spectra and the velocity shift and dispersion which together best
reproduce the observed spectrum.  Although this technique will also
fit higher-order moment terms given high-enough quality spectra,
since we are primarily interested in the velocity dispersion, and
since we are seeking to minimize the amount of averaging required to
obtain robust results, here we do not try to constrain these terms.

As described in Section~\ref{subsec:observations}, the spectral
resolution in these observations varied systematically with fibre
position on the chip.  For the derivation of the velocity shifts, this
effect is of little importance, but for measuring velocity dispersions
it could potentially introduce large systematic errors.  Thus, when
determining velocity dispersions, we take into account the position of
each fibre on the CCD.  Whenever we combine fibres from different
positions on the chip to create an averaged galaxy spectrum, we create
an associated composite stellar template by combining the stellar
spectra obtained from the same parts of the detector with the same
weighting.  This matched approach ensures that the stellar template
has the same effective instrumental resolution as the galaxy spectrum,
so that the inferred velocity dispersion is intrinsic to the galaxy,
with no artificial contribution due to a mismatch in resolution
between the galaxy and template spectra.

The errors on the derived kinematic parameters were determined using
Monte-Carlo simulations, in a similar manner to \citet{Bedregal06a}.
For each fitted spectrum, the RMS residual between the input spectrum
and the fit is determined.  We then create 100 artificial spectra by
adding random noise (with the same RMS as in the observed spectrum) to
the fitted spectrum, and measure the velocity shift and dispersion
from the simulated spectra.  The spread in the simulated values
then gives an estimate for the error in the original measurements.  In
the case of rotational velocity, the errors on the mean velocity must
be corrected for the inclination of the galaxy, and the uncertainty
in the systemic velocity of the galaxy must also be added in in
quadrature.  

Note that these random errors do not contain the full error budget, as
there may also be systematic errors arising from template mismatch,
uncertainty in the assumed inclination, systematic effects arising
from non-circular motions in the galaxy which have not been modelled,
etc.  Nonetheless, they provide a good measure of the uncertainties
that arise directly from the quality of the data.  They show, for
example, that we can obtain reasonably reliable measurements of mean
velocity and velocity dispersion from spectra with a signal-to-noise
ratio per pixel of at least 20.  

\subsection{Stellar rotation curve}
\label{subsec:rotcur}
Figure~\ref{fig:totint} illustrates the problem that we face in trying
to determine the kinematic properties of these galaxies.  It is clear
that only the very central fibres have the signal-to-noise ratios in
excess of 20 that are required to measure the velocity shift and
dispersion from the individual spectra.  For all other fibres, we need
to combine multiple spectra to increase the signal-to-noise ratio
before we can measure, for example, the velocity dispersion.

Before doing this, however, we need to correct for the rotational
motions of the stars around the centres of the galaxies.  The galaxies
in our sample are all characterised by large rotational motions, with
maximum velocities of the order of 250~\kms\ (see
Table~\ref{table:sample}), so that even spectra which have only
slightly different azimuthal angles\footnote{We define the azimuthal
angle $\phi$ as the angle with respect to the major axis measured in
the plane of the disc of the galaxy, so for a non-face-on system it
requires a deprojection of the conventional position angle.} will be
shifted to significantly different radial velocities.  Thus, even if
we simply co-added spectra close to the major axes of our galaxies, we
would artificially broaden the absorption lines in the combined
spectrum, especially when the ratio between rotational velocity and
velocity dispersion is large; for spectra further from the major axes,
the effect would be even worse.

Unfortunately, we do not know {\it a priori} exactly how fast the
stars are rotating in order to subtract this shift from the spectra
before co-adding them.  The gas rotational velocities from
\citetalias{Noordermeer07b} give an upper limit only, as the stars
will rotate more slowly due to asymmetric drift.  The magnitude of
this offset is at this stage still unknown and is in fact one of the
main quantities that we are trying to measure!

We can, however, retrieve the stellar rotation curve from the data by
realising that, when we account for the rotational motions correctly
in combining spectra, a measurement of velocity dispersion from the
combined spectrum will return a true average velocity dispersion for
the region over which spectra have been co-added.  However, if we do
not use the correct rotation velocity and consequently do not shift
the individual spectra correctly, then the resulting co-added spectrum
will always be artificially broadened and will generate a
spuriously-enlarged velocity dispersion.  Based on this effect, we
have adopted the following procedure to derive the stellar rotation
curve from these data.

First, we divide the spectra into a number of radial bins (with the
distance of each fibre to the centre of the galaxy measured in the
plane of the disc of the galaxy).  The radial extent of these bins
were chosen such that we obtain a signal-to-noise ratio of $\ga 20$ in
all the co-added spectral data that we analyze here and later, which
is sufficient to obtain a reliable measure of velocity dispersion
averaged over all the regions that we consider.  For each radial bin,
we consider a large series of possible stellar rotational velocities.
For each such velocity, we calculate the line-of-sight velocity shift
of each fibre within the bin with respect to the centre of the galaxy,
and un-Doppler-shift its spectrum accordingly.  One slight subtlety
here is that not all spectra contain the same amount of information
about the stellar rotation speed, since this motion does not project
into the line of sight along a galaxy's minor axis, so we assign a
weight to each spectrum according to its azimuthal angle, $w_s =
|\cos\phi|$.  With this weighting, we then combine all spectra in the
radial bin and measure the velocity dispersion and shift of the
resulting composite spectrum.  As discussed above, the true rotation
speed will be the one that returns the minimum measured velocity
dispersion, so we can determine the rotation speed in this bin simply
by finding the value that minimizes this quantity.

\begin{figure*}
 \centerline{\psfig{figure=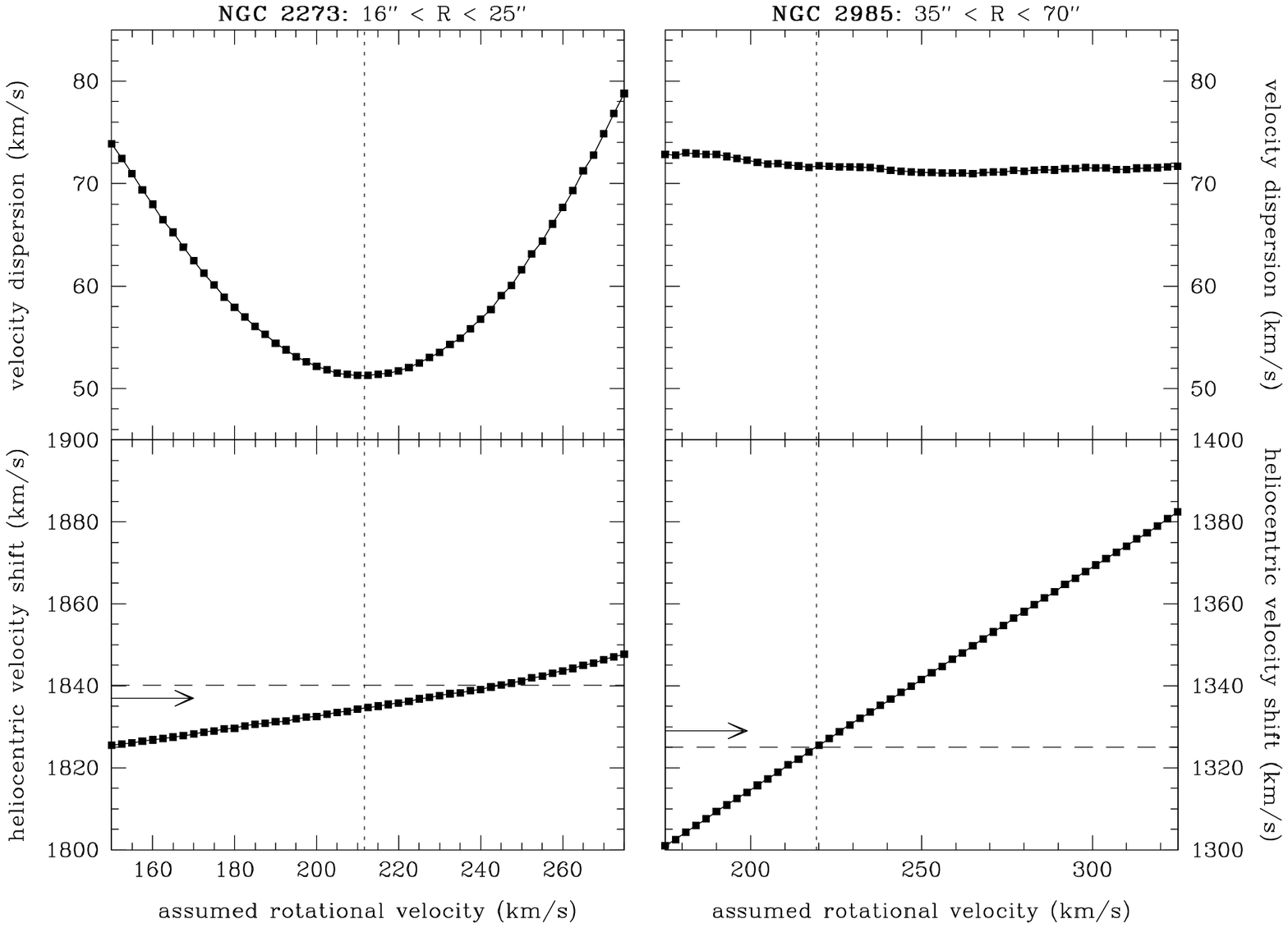,height=10.25cm}}
  \caption{Illustration of the procedure used to measure the stellar
    rotational velocities. The top left hand panel shows the measured
    velocity dispersion for the spectra at intermediate radii
    ($16\arcsec < R < 25\arcsec$) in NGC~2273, as a function of
    assumed rotational velocity. The bottom left hand panel shows the
    corresponding velocity shifts.  The right hand panels show the
    same data for the spectra at larger radii ($35\arcsec < R <
    70\arcsec$) in NGC~2985.  The vertical, dotted lines indicate the
    deduced rotational velocities. The horizontal dashed lines in the
    bottom panels indicate the derived systemic velocity of each
    galaxy; the arrows indicate the corresponding value from
    \citetalias{Noordermeer07b}.
   \label{fig:disp_vs_Vrot}} 
\end{figure*}  

This procedure is illustrated in the left-hand panels of
Figure~\ref{fig:disp_vs_Vrot}, which shows the results obtained for an
intermediate-radius bin ($16\arcsec < R < 25\arcsec$) in NGC~2273.
The curve of velocity dispersion versus assumed rotational velocity
shows a clear minimum at a rotation speed of about 212~\kms, and we
therefore conclude that this must be the true average rotation
velocity at these radii in this galaxy.

The bottom left-hand panel in Figure~\ref{fig:disp_vs_Vrot} shows that
the measured velocity shift of the composite spectrum, obtained by
optimally shifting individual spectra and averaging in azimuth, lies
very close to the systemic velocity from \citetalias{Noordermeer07b}.
This coincidence is to be expected, since, if we have adopted the
correct rotational velocity, we have effectively removed the entire
rotational component from the data and we have shifted each individual
spectrum back to the systemic velocity.  In Table~\ref{table:kindat}
we list the systemic velocities for the galaxies in the sample,
derived by taking the averages of such values from the individual
rings; the errors were estimated from the spread of the results from
the individual rings.  In each case, there is excellent agreement
between our values and the independent velocities from
\citetalias{Noordermeer07b}.  These derived systemic velocities are
also indicated in Figure~\ref{fig:disp_vs_Vrot}.

This procedure works very well for radial bins where the fibres cover
at least a nearly full ellipse around the centre of the galaxy, so that
the data probe all azimuthal angles.  However, as is apparent from
Figure~\ref{fig:totint}, we do not have such azimuthal coverage for
the outer radii in these galaxies.  The right-hand panels of
Figure~\ref{fig:disp_vs_Vrot} show what happens if we try to apply
this method at large radii where the data all lie close to the major
axis.  In these cases, all the fibres lie at similar azimuthal angles,
so are shifted by similar amounts when the rotation correction is
applied.  As a result, for any assumed rotation velocity, all spectra
are shifted by a nearly equal velocity, and the dispersion in the
resulting co-added spectrum shows very little variation (top
right-hand panel in Figure~\ref{fig:disp_vs_Vrot}).  In this case, the
method described above clearly does not have enough leverage to
determine the rotation velocity.  

%%%%%%%%%%%%%%%%%%%%%%%%%%%%%%%%%%%%%%%%%%%%%%%%%%%%%%%%%%%%%%%%%%%%%%%%%%%%%%%
%                                                                             %
% BEGIN TABLE 4: KINEMATIC PROPERTIES                                         %
% label: {table:kindat}                                                       %
%%%%%%%%%%%%%%%%%%%%%%%%%%%%%%%%%%%%%%%%%%%%%%%%%%%%%%%%%%%%%%%%%%%%%%%%%%%%%%%
\begin{table*}
 \begin{minipage}{18cm}
 \vspace{17.cm}
 \centering
  \rotcaption[Kinematic properties]
   {Measured kinematic properties. \label{table:kindat}} \hspace{6.3cm}
  \begin{rotate}{90}
   \centering
    \begin{tabular}{c r@{\hspace{0.1cm}$\pm$\hspace{0.1cm}}l l@{\hspace{1.0cm}}
       rr@{\hspace{0.1cm}$\pm$\hspace{0.1cm}}lc@{\hspace{1.0cm}}
       rr@{\hspace{0.1cm}$\pm$\hspace{0.1cm}}l
       rr@{\hspace{0.1cm}$\pm$\hspace{0.1cm}}l
       rr@{\hspace{0.1cm}$\pm$\hspace{0.1cm}}l}
    \hline
    
    galaxy & 
    \multicolumn{3}{c@{\hspace{1.0cm}}}{----------- $V_{\mathrm {sys}}$ -----------} & 
     \multicolumn{4}{c@{\hspace{1.0cm}}}{------------ stellar rotation curve ------------} & 
     \multicolumn{9}{c}{------------------ stellar velocity dispersion profiles ------------------} \\
          
           & \multicolumn{2}{c}{This study} &
     \multicolumn{1}{c@{\hspace{1.0cm}}}{\citetalias{Noordermeer07b}} &  
     \multicolumn{4}{c}{}               & 
     \multicolumn{3}{c}{major axis} & \multicolumn{3}{c}{intermediate} & 
     \multicolumn{3}{c}{minor axis} \\

           & \multicolumn{2}{c}{}          & &
     \multicolumn{1}{c}{$R_{\mathrm {eff}}$} & 
     \multicolumn{2}{c}{$V_{\mathrm {rot}}$} & 
     method & 
     \multicolumn{1}{c}{$R_{\mathrm {eff}}$} &
     \multicolumn{2}{c}{$\sigma_{\mathrm {maj}}$} & 
     \multicolumn{1}{c}{$R_{\mathrm {eff}}$} &
     \multicolumn{2}{c}{$\sigma_{\mathrm {int}}$} &  
     \multicolumn{1}{c}{$R_{\mathrm {eff}}$} &
     \multicolumn{2}{c}{$\sigma_{\mathrm {min}}$} \\

           & \multicolumn{2}{c}{\kms}      &
     \multicolumn{1}{c@{\hspace{1.0cm}}}{\kms} & 
     \multicolumn{1}{c}{\arcsec} & \multicolumn{2}{c}{\kms} & & 
     \multicolumn{1}{c}{\arcsec} & \multicolumn{2}{c}{\kms} & 
     \multicolumn{1}{c}{\arcsec} & \multicolumn{2}{c}{\kms} & 
     \multicolumn{1}{c}{\arcsec} & \multicolumn{2}{c}{\kms} \\ 
    \hline

    NGC~2273 & 1837.9 & 1.0 &    1837 &  1.8 & 104.2 & 18.8 & central fibre &  1.8 & 121.0 &  2.6 &  1.8 & 121.0 &  2.6 &  1.8 & 121.0 &  2.6 \\
             & \multicolumn{2}{c}{} & &  5.2 & 127.6 &  2.2 & minimum dispersion &  4.1 & 114.1 &  2.8 &  6.5 & 116.0 &  3.8 &  5.8 & 120.2 &  2.8 \\
             & \multicolumn{2}{c}{} & & 12.2 & 173.0 &  2.6 & minimum dispersion & 12.0 & 103.5 &  4.2 & 12.2 & 111.5 &  2.6 & 12.2 & 117.2 &  2.5 \\
             & \multicolumn{2}{c}{} & & 20.6 & 211.7 &  1.7 & minimum dispersion & 20.7 &  41.6 &  1.9 & 20.6 &  60.6 &  1.9 & 20.4 &  79.9 &  1.6 \\
             & \multicolumn{2}{c}{} & & 29.4 & 200.8 &  1.8 & minimum dispersion & 29.1 &  34.6 &  2.1 & 29.8 &  43.5 &  1.6 & 29.3 &  55.6 &  1.6 \\
             & \multicolumn{2}{c}{} & & 48.1 & 196.6 &  2.6 & minimum dispersion & 53.4 &  26.7 &  5.4 & 48.5 &  32.9 &  2.9 & 44.8 &  51.5 &  2.2 \\
             & \multicolumn{2}{c}{} & & 79.3 & 184.8 & 11.5 & velocity shift & -- & \multicolumn{2}{c}{--} & 80.5 &  21.3 & 23.8 & -- & \multicolumn{2}{c}{--} \\[0.2cm]

    NGC~2985 & 1325.0 & 4.9 &    1329 &  1.3 &  30.2 &  8.6 & central fibre &  1.3 & 141.5 &  2.0 &  1.3 & 141.5 &  2.0 &  1.3 & 141.5 &  2.0 \\
             & \multicolumn{2}{c}{} & &  3.8 & 102.4 &  8.3 & minimum dispersion &  4.7 & 136.1 &  2.3 &  3.2 & 139.3 &  1.5 &  4.5 & 133.9 &  2.5 \\
             & \multicolumn{2}{c}{} & &  8.1 & 182.2 &  8.5 & minimum dispersion &  8.9 & 126.3 &  3.3 &  7.5 & 127.8 &  2.5 &  8.2 & 117.5 &  1.8 \\
             & \multicolumn{2}{c}{} & & 13.8 & 163.4 & 13.3 & velocity shift & 14.4 & 108.4 &  3.4 & 13.4 & 105.3 &  3.1 & 13.7 & 100.1 &  1.8 \\
             & \multicolumn{2}{c}{} & & 20.4 & 200.5 & 11.3 & velocity shift & 20.6 &  77.8 &  2.8 & 20.8 & 100.1 &  3.6 & 20.1 &  97.7 &  2.6 \\
             & \multicolumn{2}{c}{} & & 29.0 & 210.1 & 10.7 & velocity shift & 29.6 &  71.5 &  3.3 & 29.0 &  87.8 &  3.0 & 28.4 &  92.4 &  3.2 \\
             & \multicolumn{2}{c}{} & & 44.7 & 219.2 & 10.5 & velocity shift & 46.4 &  70.5 &  3.8 & 43.7 &  72.6 &  4.5 & 40.8 &  95.7 &  8.1 \\
             & \multicolumn{2}{c}{} & & 62.0 & 220.2 & 10.8 & velocity shift & 62.0 &  62.3 &  4.7 & -- & \multicolumn{2}{c}{--} & -- & \multicolumn{2}{c}{--} \\
             & \multicolumn{2}{c}{} & & 79.8 & 216.1 & 12.5 & velocity shift & 79.8 &  58.8 &  6.9 & -- & \multicolumn{2}{c}{--} & -- & \multicolumn{2}{c}{--} \\
             & \multicolumn{2}{c}{} & & 99.1 & 222.9 & 31.2 & velocity shift & 99.1 &  62.1 & 23.2 & -- & \multicolumn{2}{c}{--} & -- & \multicolumn{2}{c}{--} \\[0.2cm]

    NGC~3898 & 1171.9 & 2.5 &    1172 &  1.2 &   9.3 &  3.1 & central fibre &  1.2 & 221.4 &  2.0 &  1.2 & 221.4 &  2.0 &  1.2 & 221.4 &  2.0 \\
             & \multicolumn{2}{c}{} & &  6.5 & 137.9 &  3.1 & minimum dispersion &  7.8 & 173.3 &  3.4 &  5.4 & 184.9 &  1.7 &  8.2 & 181.9 &  2.8 \\
             & \multicolumn{2}{c}{} & & 12.5 & 190.2 &  3.1 & minimum dispersion & 11.9 & 143.1 &  3.5 & 12.6 & 150.3 &  2.7 & 12.7 & 158.9 &  2.4 \\
             & \multicolumn{2}{c}{} & & 19.6 & 154.9 &  7.2 & velocity shift & 20.0 & 137.1 &  4.0 & 20.3 & 141.2 &  3.2 & 19.2 & 143.0 &  2.0 \\
             & \multicolumn{2}{c}{} & & 35.3 & 148.7 &  7.2 & velocity shift & 36.0 & 152.0 &  7.7 & 36.3 & 152.7 &  4.1 & 34.7 & 149.8 &  2.6 \\
             & \multicolumn{2}{c}{} & & 54.9 & 166.3 &  6.5 & velocity shift & 54.1 & 119.4 &  6.5 & 55.6 & 131.8 &  7.7 & 55.0 & 141.0 &  4.9 \\
             & \multicolumn{2}{c}{} & & 71.1 & 153.4 &  8.1 & velocity shift & 71.3 & 125.3 &  9.8 & 70.8 & 121.2 &  7.0 & 71.1 & 160.0 &  7.1 \\
             & \multicolumn{2}{c}{} & & 95.5 & 108.4 & 27.2 & velocity shift & -- & \multicolumn{2}{c}{--} & -- & \multicolumn{2}{c}{--} & 88.2 & 133.1 & 12.3 \\[0.2cm]

    NGC~5533 & 3847.6 & 2.2 &    3858 &  1.5 &  95.4 &  3.7 & central fibre &  1.5 & 143.4 &  2.2 &  1.5 & 143.4 &  2.2 &  1.5 & 143.4 &  2.2 \\
             & \multicolumn{2}{c}{} & &  4.6 & 184.5 &  3.2 & minimum dispersion &  3.3 & 141.1 &  1.7 &  6.0 & 137.5 &  2.7 &  5.4 & 139.4 &  2.2 \\
             & \multicolumn{2}{c}{} & & 10.9 & 224.1 &  3.7 & minimum dispersion & 10.4 & 108.7 &  2.9 & 11.1 & 121.6 &  2.6 & 11.3 & 122.7 &  3.6 \\
             & \multicolumn{2}{c}{} & & 19.8 & 241.4 &  3.7 & minimum dispersion & 19.7 & 106.6 &  3.3 & 20.1 & 112.6 &  4.1 & 19.6 & 110.4 &  3.8 \\
             & \multicolumn{2}{c}{} & & 31.6 & 219.3 &  4.3 & minimum dispersion & 31.0 & 102.0 &  4.7 & 31.7 &  92.2 &  3.3 & 32.0 & 104.5 &  5.4 \\
             & \multicolumn{2}{c}{} & & 52.1 & 208.4 &  6.2 & minimum dispersion & 55.0 &  69.4 &  8.9 & 53.4 &  54.1 &  6.5 & 46.4 &  91.2 &  8.9 \\
             & \multicolumn{2}{c}{} & & 76.5 & 174.3 & 16.8 & velocity shift & 76.4 &  56.8 & 24.4 & 76.7 &  30.5 & 19.7 & -- & \multicolumn{2}{c}{--} \\
           
    \hline
    \end{tabular} 
  \end{rotate}
  \hspace{9cm}
 \end{minipage}
\end{table*}
%%%%%%%%%%%%%%%%%%%%%%%%%%%%%%%%%%%%%%%%%%%%%%%%%%%%%%%%%%%%%%%%%%%%%%%%%%%%%%%
%                                                                             %
% END TABLE 4: KINEMATIC PROPERTIES                                           %
%                                                                             %
%%%%%%%%%%%%%%%%%%%%%%%%%%%%%%%%%%%%%%%%%%%%%%%%%%%%%%%%%%%%%%%%%%%%%%%%%%%%%%%

Nevertheless, we can still determine the rotational velocities in such
situations by considering the data in the bottom right-hand panel of
Figure~\ref{fig:disp_vs_Vrot}.  This plot shows that the velocity
shift in the co-added spectrum is a linear function of the assumed
rotation velocity.  Such a simple form is to be expected as a
consequence of the near-equal azimuth of the fibres in this bin: the
velocity shifts for all spectra in the bin are nearly equal and there
are no spectra on the opposite side of the major axis to compensate in
the final co-added spectrum.  Thus, the final spectrum has a velocity
shift directly proportional to the assumed rotational velocity and the
slope of the relation is very nearly equal to $\sin i$, where $i$ is
the inclination of the galaxy.  We can therefore identify the true
rotation velocity for this bin as that velocity for which the co-added
spectrum has a velocity shift equal to the systemic velocity of the
galaxy (as determined from the bins at smaller radii).  In effect,
this calculation is almost equivalent to simply co-adding all the
spectra, measuring a mean velocity, and subtracting off the systemic
velocity of the galaxy, but it is more accurate in that it does deal
properly with the remaining corrections due to the varying projection
into the line of sight of rotation velocity with azimuth.

Finally, the inner point in each rotation curve was derived by simply
measuring the velocity shift (with respect to the galaxy's systemic
velocity) in the fibre closest to the centre of the galaxy.  After
correcting for the inclination of the galaxy and the azimuthal angle
of the fibre, this gives a rotational velocity for the central fibre.
Note, however, that the large diameter of the fibres causes a strong
`beam smearing' effect: due to the large velocity gradients in the
centres of these galaxies, the central fibre probes stars with different
velocities and the measured velocity shift is the result of a
complicated convolution of the stellar velocity field and the
brightness distribution.  Thus, the central points in the rotation
curves should only be used with caution.
\begin{figure*}
 \centerline{\psfig{figure=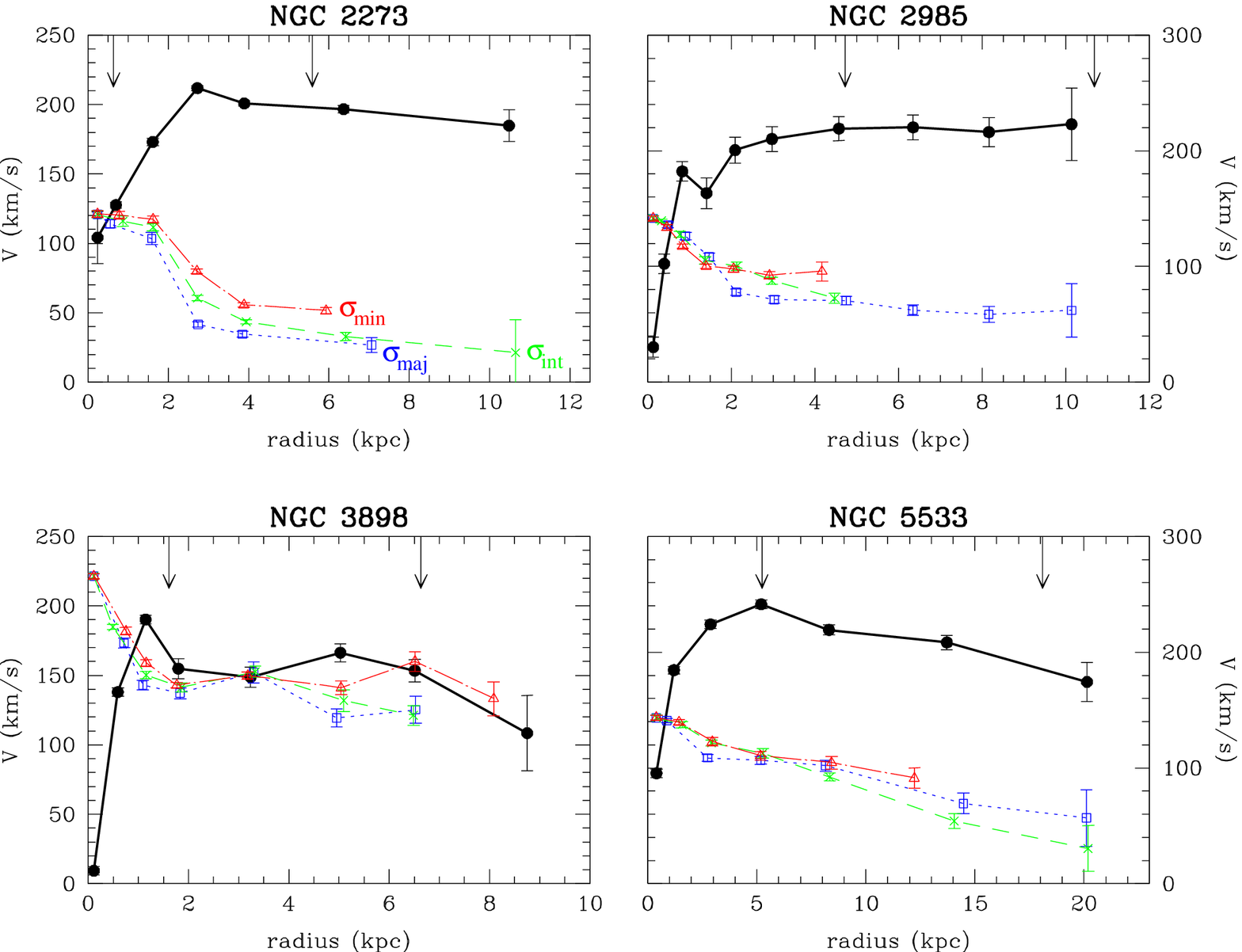,height=11cm}}
  \caption{Stellar rotation curves and velocity dispersion
  profiles. Bold lines and bullets show the stellar rotation
  curves. Squares connected by dotted lines, crosses connected by
  dashed lines and triangles connected by dot-dashed lines indicate
  the velocity dispersions along the major, intermediate and minor
  axes respectively. Error bars show the formal measurement errors
  (see Section~\ref{subsec:PPXF} for details). The inner vertical
  arrows indicate the transition radius between bulge and disc
  dominated photometry, while the outer arrows indicate two disc scale
  lengths. \label{fig:kinprofs}}
\end{figure*}

The resulting rotation curves for the four galaxies in the sample are
tabulated in Table~\ref{table:kindat} and plotted in black in
Figure~\ref{fig:kinprofs}.  The quoted radii are the
luminosity-weighted average radii of all fibres in each bin.

\subsection{Stellar velocity dispersions}
\label{subsec:veldisp}
The method described above to derive the stellar rotation curves also
yields an average velocity dispersion for each radial ring.  However,
since the velocity dispersion tensor can be quite strongly
anisotropic, and different azimuth angles give different projections
of this tensor, we might expect the measured line-of-sight velocity
dispersion to vary systematically with azimuth.  To quantify this
effect, we divided each radial bin in three sub-bins, according to the
azimuthal angles of the fibres.  For each radial bin, we combine all
spectra within 30\deg\ of the major axis, between 30 and 60\deg\
from the major axis, and within 30\deg\ of the minor axis (all
measured in the plane of the galaxy disc), as shown in
Figure~\ref{fig:totint}.  Within each sub-bin, the individual spectra
are shifted by an amount that subtracts out the rotational velocity
derived above, as appropriate for that individual fibre's azimuthal
angle.  The shifted spectra in the sub-bin are then co-added with no
weighting, and the velocity dispersion derived for the composite
spectrum.

The resulting profiles of velocity dispersion along the major,
intermediate and minor axes are listed in Table~\ref{table:kindat} and
plotted in Figure~\ref{fig:kinprofs}.  The radii are, again, the
luminosity weighted average radii of all fibres in each sub-bin; the
small differences between the effective radii for the three sub-bins
in a given radial bin reflect the details of the fibre locations and
brightness distribution of each galaxy.  Generally speaking, these
velocity dispersion profiles are well behaved in the sense that they
vary smoothly with radius, and show a systematic ordering in amplitude
between minor, intermediate and major axes, characteristic of the
detection of an anisotropic velocity dispersion.  The orderliness of
these data must to some extent be due to the particularly simple
featureless early-type disks selected for this sample, but it also
reflects the care taken in obtaining the data and analyzing it in an
optimal manner.

%%%%%%%%%%%%%%%%%%%%%%%%%%%%%%%%%%%%%%%%%%%%%%%%%%%%%%%%%%%%%%%%%%%%%%%%%%%%%%%
%                                                                             %
%  4. Modeling the dynamics                                                   %
%  \label{sec:dynamics}                                                       %
%                                                                             %
%%%%%%%%%%%%%%%%%%%%%%%%%%%%%%%%%%%%%%%%%%%%%%%%%%%%%%%%%%%%%%%%%%%%%%%%%%%%%%%
\section{Modelling the dynamics}
\label{sec:dynamics}
\begin{figure*}
 \centerline{\psfig{figure=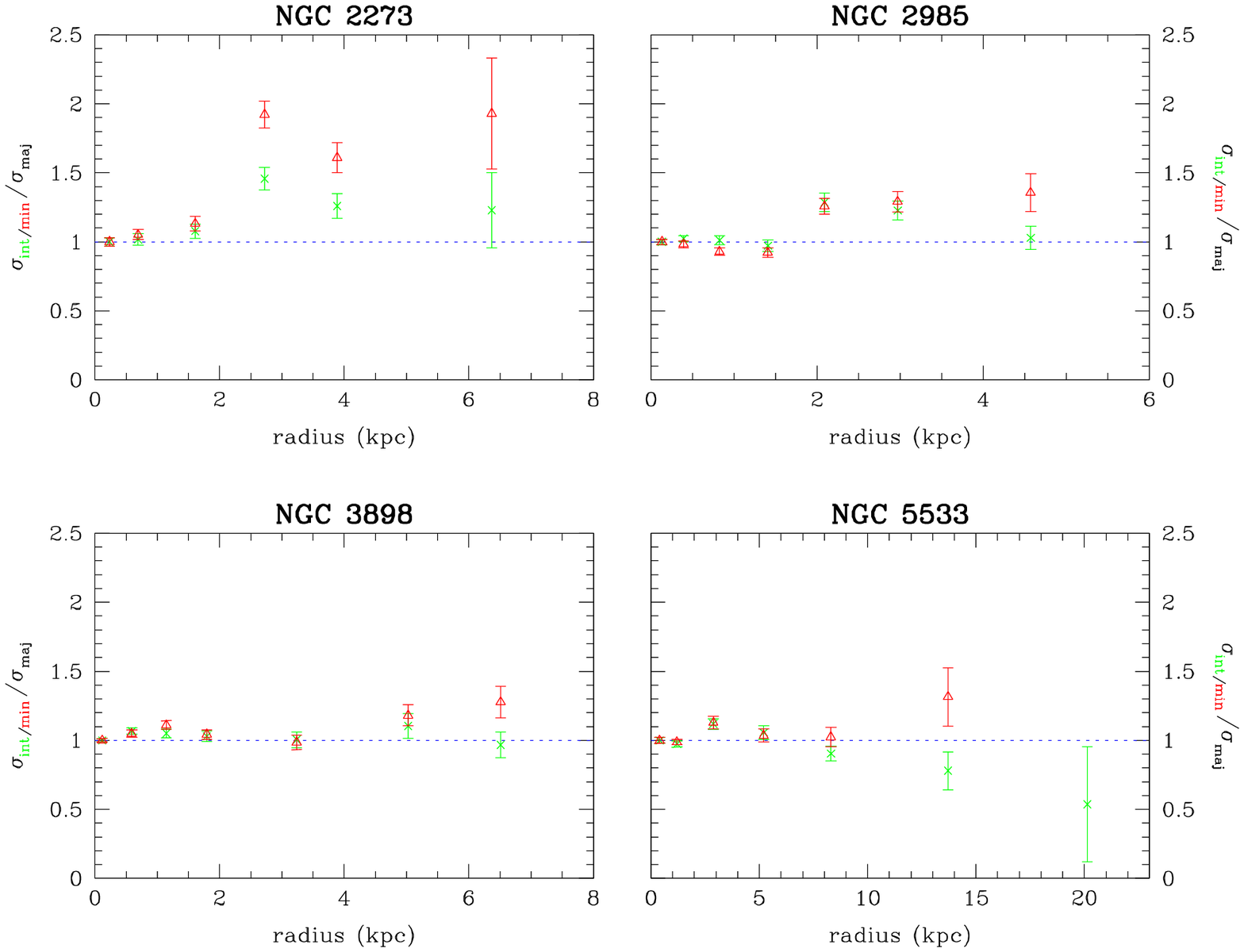,height=11cm}}
  \caption{Ratios of observed velocity dispersions as a function of
    radius. The red triangles and green crosses show, respectively,
    the ratio of the velocity dispersions along the minor
    and intermediate axes to those on the major
    axis. \label{fig:ellipsoid}}
\end{figure*}

Having obtained these high-quality kinematic data, we are now in a
position to begin to interpret them both qualitatively and
quantitatively in terms of the galaxies' stellar dynamics.  An
interesting qualitative diagnostic comes from plotting the ratio of
velocity dispersions along major and minor axes versus radius.  As
Figure~\ref{fig:ellipsoid} shows, this plot confirms that the velocity
dispersions show an orderly variation with radius.  Inclusion of the
ratio of the intermediate-angle velocity dispersion to the major axis
dispersion on this plot further confirms the orderliness of the data.
For NGC~3898 and NGC~5533, the data show no evidence that these ratios
depart from unity, consistent with an isotropic velocity dispersion
tensor.  This simplicity could well be associated with the ``hotness''
of the disks in these systems, particularly in the case of NGC~3898
where random motions remain comparable to rotational ones at all
radii: as indicated by \citet{Merrifield01} and \citet{Shapiro03},
earlier-type galaxies, which typically have hotter disks, also tend to
have simpler more isotropic velocity dispersions.

The behaviour for NGC~2273 and NGC~2985 is more interesting.  At small
radii, these systems display the isotropic velocity dispersions that
one would expect for the centre of a hot-bulge stellar population.
However, they undergo what appears to be a quite sudden transition to
a different regime in which the minor axis dispersion exceeds the
major axis one.  Since the minor-axis line-of-sight velocity
dispersion contains a large component of the radial intrinsic velocity
dispersion, while the major-axis line-of-sight velocity dispersion is
dominated by the azimuthal intrinsic velocity dispersion, this
departure from unity would indicate that in the outer regions of these
galaxies the velocity dispersion tensor is radially anisotropic.

Such anisotropy is expected in a colder disk population, so this
transition seems a fairly clear indicator of a transition from the
bulge to disk in these systems.  However, comparison with
Figure~\ref{fig:kinprofs} shows that any such naive decomposition into
bulge and disk should be treated with some caution, as the radius at
which this transition in the velocity dispersion tensor occurs does
not correspond to the radius at which the photometric decomposition
indicates a transition from bulge to disk, nor does it occur near the
radius at which rotational motion starts to dominate over random
motions.  The lack of correspondence between the kinematic and
photometric decompositions provides a health warning about the lack of
uniqueness in the photometric approach, reflecting the loss of
information when the full dynamical phase space is not accessible.
The lack of correspondence between the transition to disk-like
kinematics and the radius at which rotational motion dominates is more
interesting, and presumably reflects the more complex nature of the
internal dynamics of these systems.  It is, however, notable that the
transition in the velocity dispersions does correspond closely to the
radius at which the rotational motion flattens out to its
approximately-constant asymptotic value, which is also the regime in
which one would expect the dynamics to take on a relatively simple
disk-like form.

\subsection{Asymmetric drift}
\label{subsec:asymmetricdrift}
We now turn to more quantitative consideration of the dynamics of
these systems, specifically seeking to understand whether the stellar
motions respect the asymmetric drift equation, as we might expect.
This equation, which quantifies the lag due to random stellar motions
between mean rotational motion and the local circular speed, can be
written
\begin{equation}
  V_{\mathrm c}^2 = V_{\mathrm {rot}}^2 + \sigma_\phi^2 - 
          \sigma_R^2 \left( 1 + \frac{d (\ln \nu)}{d (\ln R)} \right) -
          R \frac{d \sigma_R^2}{d R}.
  \label{eq:asymdrift}
\end{equation}
where $V_{\mathrm c}$ is the local circular speed, $V_{\mathrm {rot}}$
is the measured rotational velocity of the stars at this radius,
$\sigma_\phi$ and $\sigma_R$ are the azimuthal and radial components
of the velocity dispersion ellipsoid and $\nu$ is the stellar number
density \citep{Binney87}.  In this formulation, we have followed the
usual procedure of neglecting the tilt term, $d (\overline{V_RV_z}) /
d z$, which is assumed to be small compared to the other terms.

\begin{figure*}
 \centerline{\psfig{figure=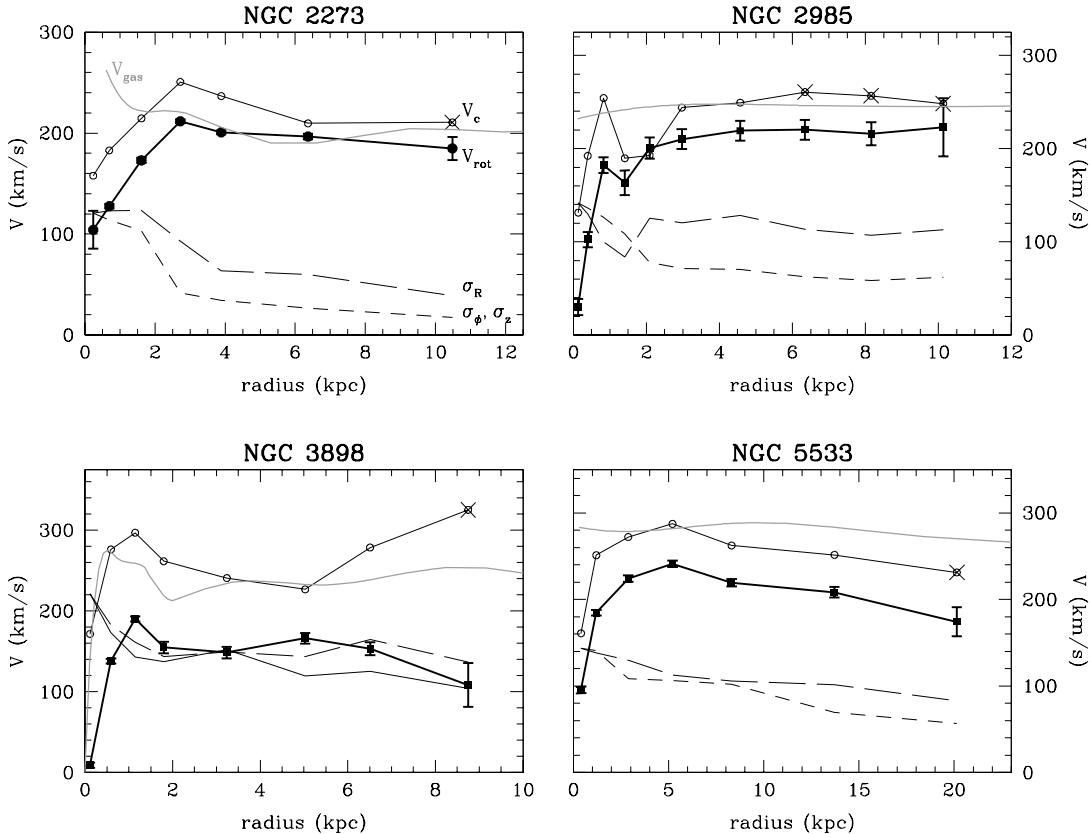,height=11cm}}
  \caption{Derived dynamical properties of the sample galaxies. Bold
    lines and bullets show the raw rotational velocities (as in
    Figure~\ref{fig:kinprofs}), while thin solid lines and open
    symbols show the stellar rotation curves after correction for
    asymmetric drift using equations~\ref{eq:asymdrift} and
    \ref{eq:sigma_phi&R}.  Crosses indicate points where the velocity
    dispersions were not measured along all three axes. Grey lines
    show the gas rotation curves from
    \citetalias{Noordermeer07b}. Short and long dashed lines show,
    respectively, the azimuthal and radial components of the velocity
    dispersion ellipsoid. \label{fig:asymdrift}}
\end{figure*}

The intrinsic components of velocity dispersion, $\sigma_\phi$ and
$\sigma_R$, are related to the observed major- and minor-axis velocity
dispersions by the relations
\begin{eqnarray}
  \label{eq:ellipsoid}
  \sigma_{\mathrm {maj}}^2 & = & \sigma_\phi^2 \, \sin^2 \! i + 
                                 \sigma_z^2 \, \cos^2 \! i \nonumber \\[-0.2cm]
                           &   &                                     \\[-0.2cm]
  \sigma_{\mathrm {min}}^2 & = & \sigma_R^2 \, \sin^2 \! i + 
                                    \sigma_z^2 \, \cos^2 \! i,   \nonumber
\end{eqnarray}
which also include the contributions from the velocity dispersion in
the third, vertical, direction, $\sigma_z$.  Without additional
information, one cannot derive all three components of the velocity
dispersion from these two observed quantities.  However, three of the
four galaxies in this sample are sufficiently close to edge-on that
very little of the vertical motion projects into the line of sight (we
return to the fourth, NGC~2985, below).  For these galaxies, the
assumed behaviour of this third component has very little impact on
the inferred values for the other two components.  For simplicity, and
motivated by the apparent isotropy in two of these galaxies, we
therefore make the simplest possible assumption that $\sigma_z =
\sigma_\phi$.  Under this assumption, equation~\ref{eq:ellipsoid}
reduces to:
\begin{eqnarray}
  \label{eq:sigma_phi&R}
  \sigma_\phi & = & \sigma_{\mathrm {maj}} \nonumber    \\
              &   &                                     \\[-0.4cm]
  \sigma_R    & = & \sqrt{ \frac{\sigma_{\mathrm {min}}^2 - 
                                 \sigma_{\mathrm {maj}}^2 \cos^2 \! i }
                                { \sin^2 \! i} }. \nonumber
\end{eqnarray}
The profiles for $\sigma_\phi$ and $\sigma_R$ obtained from this
equation are shown in Figure~\ref{fig:asymdrift}.  For radii where we
have not measured $\sigma_{\mathrm {maj}}$ or $\sigma_{\mathrm {min}}$
(but where we have at least one of $\sigma_{\mathrm {maj}}$,
$\sigma_{\mathrm {int}}$ or $\sigma_{\mathrm {min}}$), we use the fact
that the ratios of these quantities as shown in
Figure~\ref{fig:ellipsoid} are consistent with remaining constant at
large radii, in order to extrapolate values.  

Equation~\ref{eq:asymdrift} also requires the density of the stellar
tracer, $\nu(R)$, which we obtain directly from the R-band photometric
profiles from \citet{Noordermeer07a}.  Such red light traces the
stellar population whose kinematics we are measuring here quite
closely, although for such early-type disk galaxies, which contain
little by way of colour gradients, the choice of band is not critical.
More challengingly, equation~\ref{eq:asymdrift} also demands a measure
of the radial gradient in the velocity dispersion.  Motivated by a
number of previous observations \citep[e.g.][]{VanderKruit86,
Lewis89, Bottema93}, this derivative has conventionally
been obtained by assuming that the profile declines exponentially
\citep{Gerssen97, Westfall08}.  However, it is apparent from
Figure~\ref{fig:asymdrift} that these higher-quality data are really
not consistent with such a simple assumed form -- a point also
uncovered in other recent studies of velocity dispersion profiles in
disk galaxies \citep[e.g.][]{Merrett06, Noordermeer08}.  Accordingly,
we have adopted the less parametric approach of interpolating the
velocity dispersion profile on to a polynomial and calculating its
derivative numerically.  Such a procedure will tend to amplify any
noise in the data, but such noise amplification represents the true
uncertainty that arises from the presence of this derivative in the
equation, so is an honest and unbiased way to include this term in the
calculation.

Inserting all these ingredients into equation~\ref{eq:asymdrift}
allows us to solve for the circular velocity as a function of radius
in these galaxies, $V_c(R)$, as shown in Figure~\ref{fig:asymdrift}.
As described above, we can compare these inferred values to the
rotation curves obtained directly from emission-line data, which are
also shown in Figure~\ref{fig:asymdrift}.  On the whole, the derived
rotation curves match reasonably well, in that the correction for
asymmetric drift shifts the stellar streaming data significantly
closer to the true rotation curve, and in several cases provides an
almost-exact match.  The only real exception is NGC~2273, where the
asymmetric drift correction over-enhances the inferred rotation curve
modestly above its true value.  It is notable that this is by far the
coldest disk system in the sample, and as such it is most likely to
fall victim to any residual systematic errors that will tend to
over-estimate the velocity dispersions and hence over-correct for
asymmetric drift.  However, for such a system the asymmetric drift
correction is in any case small, so the associated small bias does not
have a large impact on the inferred rotation curve.  The last point in
the NGC~3898 data is also high, but the error bars are large, and the
derivative term has a very large uncertainty without bracketing data
from which to interpolate.  Even when the asymmetric drift term is
large, both in the bulge-dominated regions of all the systems and the
outer parts of those with hot disks, the inferred rotation curve
matches the true one reasonably well, which bodes well for the
application of such corrections to generally-hot S0 systems that lack
the external measure of the rotation curve afforded by emission-line
data.

Note that the ansatz that $\sigma_z = \sigma_\phi$ had no real
physical basis.  In fact, it would be dynamically more interesting to
investigate whether the data are consistent with $\sigma_z =
\sigma_R$, as such an equality is required if the underlying
distribution function respects only two integrals of motion
\citep{Jeans15}.  As described above, most of the galaxies are
sufficiently far from face-on that different assumptions about
$\sigma_z$ make little difference to the observables.  However,
NGC~2985 is close enough to face-on for a significant amount of
motions perpendicular to the plane to project into the line of sight,
so for this galaxy we can investigate the behaviour of $\sigma_z$, and
hence test this possibility.    

\begin{figure}
 \centerline{\psfig{figure=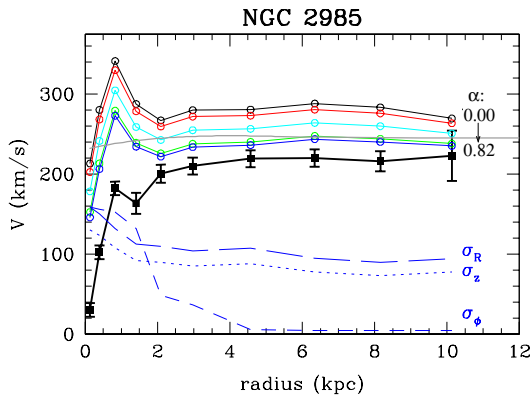,
     height=5.15cm}}
  \caption{The effect of varying assumptions for the vertical velocity
    dispersion on the corrections for asymmetric drift in
    NGC~2985. The bold line and bullets show the raw rotational
    velocities of the stars (as in Figure~\ref{fig:kinprofs}) and the
    gray line shows the gaseous rotation curve from
    \citetalias{Noordermeer07b}. The thin black, red, cyan, green and
    blue lines and open circles show the rotation curves, corrected
    for asymmetric drift assuming ratios of vertical-to-radial
    velocity dispersion of respectively $\alpha=0$, 0.25, 0.5, 0.75
    and 0.82. The blue dotted, long and short dashed lines show the
    profiles of the vertical, radial and tangential velocity
    dispersion for the limiting case of
    $\alpha=0.82$. \label{fig:asymdrift_vs_sigmaz}}
\end{figure}

Figure~\ref{fig:asymdrift_vs_sigmaz} shows the corrected rotation
curves for a range of different values adopted for the ratio $\alpha =
\sigma_z / \sigma_R$, ranging from $\alpha = 0$ to $\alpha = 0.82$.
It turns out that $\alpha=0.82$ is the maximum possible value, since
at this value $\sigma_z$ is responsible for all the observed velocity
dispersion along the major axis, so $\sigma_\phi$ goes to zero (see
Figure~\ref{fig:asymdrift_vs_sigmaz}).  Any larger value for $\alpha$
would further increase $\sigma_z$, making $\sigma_\phi$ unphysically
imaginary (see equation~\ref{eq:ellipsoid}).  Thus, we can robustly
rule out the possibility that $\sigma_z = \sigma_R$ in this system:
perhaps unsurprisingly, the disk distribution function in NGC~2985
requires a third integral.  

Clearly, even pushing $\alpha$ as high as 0.82 results in a physically
implausible model with zero velocity dispersion in the tangential
direction.  It also conflicts with the observations in that, as
Figure~\ref{fig:asymdrift_vs_sigmaz} shows, it fails to reproduce the
emission-line-derived rotation curve.  From this additional
constraint, one can see that for a model to match this observation we
require a value of $\sigma_z / \sigma_R$ of approximately 0.7.  In
fact, the ratio is essentially identical to the model shown in
Figure~\ref{fig:asymdrift}, since for this new model it also turns out
that $\sigma_z \approx \sigma_\phi$, as previously assumed.  By using
this additional constraint of requiring the observed asymmetric drift
to be reproduced, as well as the two observed components of velocity
dispersion, we have effectively closed the system to show that the
velocity ellipsoid in the disk of this system, if its shape is
approximately constant with radius, must have principal axes in the
ratio $\sigma_z : \sigma_\phi : \sigma_R \approx 0.7 : 0.7 : 1$.

This ratio is instructive because the rotation curve for NGC~2985 is
very close to being flat, for which the epicyclic approximation would
also predict a value of $\sigma_\phi/\sigma_R = 1/\sqrt{2} \approx
0.7$, quite independent of the above analysis.  At least for this
system, it would appear that the concerns over the validity of the
epicyclic approximation are unfounded, providing more confidence in
its use for systems even where the ratio of ordered-to-random
velocities is as low as $\sim 3$.

\section{Conclusions}
\label{sec:conclusions}
In this paper, we have presented deep IFU observations of four nearby
early-type disk galaxies with good emission-line rotation curves.
These high-quality data were obtained with the motivation both of
developing new techniques for the optimal extraction of kinematic
parameters, and of testing the frequently-adopted dynamical formulae
of the asymmetric drift equation and the epicyclic approximation.  The
conclusions of this study are as follows:
\begin{itemize}
\item With care, it is possible to extract the rather subtle dynamical
  signature of a disk system from two-dimensional spectral data.  In
  particular, one can avoid the biases produced by the artificial
  enhancement of velocity dispersion that occurs when data are
  inappropriately averaged together.
\item With such data, application of the asymmetric drift equation
  seems to work reasonably well, in that one can reproduce the
  rotation curve as inferred independently from emission-line data.
  There is a caveat that residual systematic errors may compromise the
  results to some extent for particularly cold systems, but for such
  systems the asymmetric drift corrections are small, so the
  corresponding small bias is not too debilitating.  For hotter
  systems, the inferred accuracy of the correction is important for
  the credibility of the rotation speeds inferred for S0 galaxies in
  studies of their Tully--Fisher relation, as one does not usually
  have the luxury of an emission-line rotation curve in such systems.
\item For an appropriately-inclined system like NGC~2985, one can
  obtain a strong upper limit on the ratio of $\sigma_z/\sigma_R$,
  which rules out the possibility that the system only respects two
  integrals of motion.
\item Such data, combined with the asymmetric drift equation, allow
  one to solve for the full three-dimensional shape of the velocity
  dispersion tensor.  In the case of NGC~2985, even though it is a
  relatively-hot early-type disk, the resulting ratio of velocity
  dispersions is quite consistent with the prediction of simple
  epicyclic theory, giving some further confidence in the general
  applicability of this approximation.
\end{itemize}
Clearly, with the development of large-field high-dispersion IFUs
like PPAK, it is now possible to map out the kinematics of nearby disk
galaxies in great detail, and hopefully this paper has demonstrated
that it is possible to analyze such data and model them using the
appropriate dynamical equations, with the aim of understanding their
full phase-space structure, and, ultimately, their formation.

\section*{Acknowledgements}
We are greatly indebted to Marc Verheijen, for his help in the
preparation of the observations as well as the data reduction, and
Alejandro Bedregal for his advice on the analysis and interpretation
of the spectra.  We would also like to thank the staff at the Calar
Alto Observatory for their assistance during the observations and the
data reduction.  This work is based on observations collected at the
Centro Astron\'omico Hispano Alem\'an (CAHA) at Calar Alto, operated
jointly by the Max-Planck Institut f\"ur Astronomie and the Instituto
de Astrof\'isica de Andaluc\'ia (CSIC).  The observations were funded
by the Optical Infrared Coordination Network (OPTICON), a major
international collaboration supported by the Research Infrastructures
Programme of the European Commission's Sixth Framework Programme.  MRM
gratefully acknowledges the support of a PPARC/STFC Senior Fellowship.

\end{document}